\documentclass[12pt]{article}

\catcode`\@=11
\global\arraycolsep=2pt
\usepackage{amsbsy,amssymb,latexsym,amsfonts,amsmath}
\usepackage{graphicx,color}
\usepackage[numbers]{natbib}

\usepackage{hyperref}
\hypersetup{
    colorlinks=true,  
    linkcolor=blue,  
    citecolor=blue,   
    urlcolor=blue    
}

\begin{document}

\begin{flushright}
YITP-26-13
\end{flushright}

\vspace*{0.7cm}

\begin{center}
{ \Large  To boost or not to boost, that's the question}
\vspace*{1.5cm}\\
{Yu Nakayama}
\end{center}
\vspace*{1.0cm}
\begin{center}

Yukawa Institute for Theoretical Physics,
Kyoto University, Kitashirakawa Oiwakecho, Sakyo-ku, Kyoto 606-8502, Japan

\vspace{3.8cm}
\end{center}

\begin{abstract}
Or should we talk about dS/CFT correspondence or dS/SFT correspondence in cosmological correlators?  In non-unitary field theories---which are conjectured to be dual to cosmological correlators---scale invariance does not necessarily imply full conformal invariance. While general relativity predicts the emergence of conformal invariance (or boost symmetry in the bulk), various modified theories of gravity suggest only scale invariance, characterized by the absence of bulk boost symmetry. We demonstrate this distinction using Einstein-Aether theory as a canonical example.
\end{abstract}

\setcounter{page}{0}

\newpage

\section{Introduction}

Observationally, our universe selects a preferred reference frame---the rest frame of the cosmic microwave background. While this implies a broken boost symmetry in the cosmological background, the fundamental laws of particle physics have preserved Lorentz invariance so far in any experiment. Does this preferred frame originate merely from the spontaneous breaking by the background solution, or is it embedded within the fundamental principles of gravity itself? Indeed, the distinction between initial conditions and the fundamental constituents of the theory becomes subtle when addressing the origin of the universe.

While general relativity is formulated as a diffeomorphism-invariant theory with no preferred direction, many modified theories of gravity introduce a dynamical preferred frame at the fundamental level~\cite{Jacobson:2000xp}. It is plausible that the Lorentz invariance observed in the Standard Model is an emergent or approximate symmetry, potentially violated in the far IR (cosmological scales) or the far UV (quantum gravity scales)~\cite{Horava:2009uw, Blas:2010hb}. See e.g. \cite{Mattingly:2005re,Jacobson:2007veq,Mukohyama:2010xz,Liberati:2013xla,Wang:2017brl,Will:2014kxa} for reviews. 

In the context of quantum gravity, the holographic principle offers a powerful framework to scrutinize these symmetries. In a cosmological setting, assuming a holographic dual exists, the question of bulk symmetries translates into properties of the holographic correlation functions. In holographic field theory, the corresponding question arises naturally: does scale invariance of the dual field theory necessarily imply invariance under special conformal transformations?

We focus on an exponentially expanding universe, approximating de Sitter spacetime. In the holographic dictionary, the dilatation symmetry of the holographic field theory corresponds to the isometry of the bulk de Sitter spacetime~\cite{Strominger:2001pn, Maldacena:2002vr}. Among others, the special conformal transformations on the holographic field theory are dual to the spatial boosts in the bulk. Thus, checking for conformal invariance in the holographic field theory is equivalent to testing the boost symmetry of the bulk gravitational theory.

In standard unitary quantum field theories, scale invariance typically implies full conformal invariance. This enhancement occurs because, for a theory to be scale invariant but not conformal invariant, the trace of the stress tensor must be a total divergence of a ``virial current''---a condition that is severely constrained when unitarity (or reflection positivity in the Euclidean signature) is imposed. Indeed, following the seminal ideas of Zamolodchikov~\cite{Zamolodchikov:1986gt} and earlier work by Mack~\cite{Mack:1975je}, Polchinski proved this equivalence as a theorem in two dimensions~\cite{Polchinski:1987dy}. Significant progress regarding this enhancement in higher dimensions has also been reported more recently in~\cite{Dorigoni:2009ra,Luty:2012ww,Fortin:2012hn,Dymarsky:2013pqa}. See \cite{Nakayama:2013is} for a review.

If this theorem were directly applicable to cosmology, a scale-invariant universe would inevitably possess full de Sitter isometries with boost symmetry. However, a pivotal distinction lies in the nature of the dual theory: in cosmological holography (such as dS/CFT), the holographic theory is Euclidean and non-unitary (or not reflection positive)~\cite{McFadden:2009fg, Anninos:2011ui,Nakata:2020luh}. Once the unitarity constraint is relaxed, the implication ``scale $\to$ conformal" no longer holds. The virial current can be non-zero, signaling a violation of special conformal symmetry (and hence bulk boosts) without breaking scale invariance.\footnote{The discussions of scale vs conformal in the context of asymptotic AdS holography can be found in various studies, including~\cite{Nakayama:2009qu,Nakayama:2009fe,Nakayama:2011zw,Nakayama:2012sn,Nakayama:2016ydc,Nakayama:2016xzs,Chawla:2025fpn}. The consensus suggests that such a realization is challenging unless unitarity is abandoned.}

In this paper, we demonstrate how this mechanism is realized in boost-breaking cosmologies. We employ the Einstein-Aether theory~\cite{Jacobson:2000xp} as a canonical framework to explicitly construct scale-invariant holographic correlators that lack full conformal symmetry. Our analysis reveals that the subleading mode of the aether field in the bulk generates a non-vanishing virial current in the holographic field theory. While we focus on the Einstein-Aether model, our results capture generic features relevant to other modified gravity theories, such as Ho\v{r}ava gravity~\cite{Horava:2009uw} or shift-symmetric Horndeski theories~\cite{Deffayet:2011gz,Kobayashi:2011nu,Jacobson:2010mx,Audren:2013dwa} including the ghost condensation~\cite{Arkani-Hamed:2003pdi,Mukohyama:2006be}.\footnote{When the shift symmetry is gauged, we naturally obtain the vector-tensor theory~\cite{Nakayama:2009qu,Aoki:2021wew}, whose low-energy limit is typically given by the Einstein-Aether theory.}

In recent years, there has been significant interest in cosmological correlation functions from the bootstrap method~\cite{Baumann:2022jpr}. There, the generalizations of the cosmological bootstrap with the boost symmetry breaking are known as the boostless bootstrap~\cite{Pajer:2020wnj,Pajer:2020wxk,Baumann:2020dch,Stefanyszyn:2020kay,Jazayeri:2021fvk,Pimentel:2022fsc,Bonifacio:2022vwa,Ghosh:2023agt,Du:2024hol}. In these studies, the breaking of the boost symmetries is often accompanied by the breaking of the cosmological time dilatation symmetry (or scale symmetry). In this context, our analysis can be interpreted as a special situation when we only allow the violation of the boost symmetry without the violation of the time dilatation.

The organization of the paper is as follows. In section 2, we introduce the Einstein-Aether theory and the background de Sitter solution without the boost symmetry. In section 3, we show how the particular structure of the holographic stress tensor for scale invariance without conformal invariance is realized in the cosmological solutions. In section 4, we present an example of holographic computations. In section 5, we conclude with some discussions.

\section{Einstein-Aether Theory}
\subsection{Action and Equations of Motion}
We consider a four-dimensional spacetime with coordinates $x^\mu$ and metric signature $(-,+,+,+)$. The Einstein-Aether theory is defined by the Einstein-Hilbert action coupled to a dynamical, unit-norm timelike vector field $u^\mu$. The total action $S$ is given by~\cite{Jacobson:2000xp, Jacobson:2007veq}:
\begin{equation}
    S = \int d^4x \sqrt{-g} \left[ M_{\mathrm{pl}}^2 (R - 2\Lambda) + \mathcal{L}_{\mathrm{AE}} \right].
\end{equation}
The Lagrangian density for the aether field is constructed as follows:
\begin{equation}
    \mathcal{L}_{\mathrm{AE}} = -M_{\mathrm{pl}}^2 \left[ K^{\alpha\beta}_{\mu\nu} (\nabla_\alpha u^\mu) (\nabla_\beta u^\nu) - \lambda(g_{\mu\nu} u^\mu u^\nu + 1) \right].
\end{equation}
Here, we have factored out $M_{\mathrm{pl}}^2$ from the entire aether sector for convenience. $\lambda$ is a Lagrange multiplier field that enforces the unit-norm constraint $u^2 = -1$.
The tensor $K^{\alpha\beta}_{\mu\nu}$ defining the kinetic terms is given by:
\begin{equation}
    K^{\alpha\beta}_{\mu\nu} \equiv c_1 g^{\alpha\beta} g_{\mu\nu} + c_2 \delta^\alpha_\mu \delta^\beta_\nu + c_3 \delta^\alpha_\nu \delta^\beta_\mu - c_4 u^\alpha u^\beta g_{\mu\nu}.
\end{equation}
The dimensionless parameters $c_i$ determine the dynamics of the aether field.  Our convention aligns with  \cite{Eling:2014saa,Armendariz-Picon:2010aer}.\footnote{Note that the parameters $c_i$ used in \cite{Armendariz-Picon:2010aer} are defined with opposite signs to those in \cite{Eling:2014saa}. We follow the convention used in \cite{Eling:2014saa} (except for the sign of the Lagrange multiplier $\lambda$) in agreement with the one used in \cite{Jacobson:2007veq}.}

By varying the action with respect to $u^\mu$ and $\lambda$, we obtain the equations of motion. The variation with respect to $\lambda$ yields the constraint $u^2 = -1$. The equation of motion for the aether field $u^\mu$ is given by~\cite{Eling:2003rd, Eling:2006aw,Eling:2014saa}:
\begin{equation}
    \label{eq:aether_EOM}
    \nabla_\alpha J^{\alpha}_{\phantom{\alpha}\mu} + c_4 a_\alpha \nabla_\mu u^\alpha + \lambda u_\mu = 0,
\end{equation}
where the aether current is $J^{\alpha}_{\phantom{\alpha}\mu} \equiv K^{\alpha\beta}_{\mu\nu} \nabla_\beta u^\nu$, and the acceleration is $a^\mu \equiv u^\nu \nabla_\nu u^\mu$.

Contracting \eqref{eq:aether_EOM} with $u^\mu$ determines the Lagrange multiplier $\lambda$:
\begin{equation}
    \lambda = u^\mu \nabla_\alpha J^{\alpha}_{\phantom{\alpha}\mu} + c_4 a^2.
\end{equation}

Finally, varying with respect to $g^{\mu\nu}$ gives the Einstein equations:
\begin{equation}
    G_{\mu\nu} + \Lambda g_{\mu\nu} = M_{\mathrm{pl}}^{-2} T_{\mu\nu}^{\mathrm{AE}}.
\end{equation}
The energy-momentum tensor of the aether field takes the form~\cite{Jacobson:2004ts, Eling:2006aw}:
\begin{align}
    T_{\mu\nu}^{\mathrm{AE}} &= M_{\mathrm{pl}}^2 \bigg\{ \nabla_\alpha \left( J^\alpha_{\phantom{\alpha}(\mu} u_{\nu)} - J_{(\mu}^{\phantom{(\mu}\alpha} u_{\nu)} + J_{(\mu\nu)} u^\alpha \right) \nonumber \\
    &\quad + c_1 \left[ (\nabla_\mu u_\alpha)(\nabla_\nu u^\alpha) - (\nabla_\alpha u_\mu)(\nabla^\alpha u_\nu) \right] \nonumber \\
    &\quad + c_4 a_\mu a_\nu - \lambda u_\mu u_\nu \bigg\} + \frac{1}{2} g_{\mu\nu} \mathcal{L}_{\mathrm{AE}}^{\text{kin}},
\end{align}
where $\mathcal{L}_{\mathrm{AE}}^{\text{kin}} \equiv -M_{\mathrm{pl}}^2 K^{\alpha\beta}_{\mu\nu} (\nabla_\alpha u^\mu) (\nabla_\beta u^\nu)$ is the kinetic term in the aether action.

\subsection{Cosmological Background and Symmetry}

We begin by establishing that the Einstein-Aether theory admits a de Sitter solution. Let us consider the Poincar\'e patch of de Sitter spacetime, described by the metric:
\begin{equation}
    ds^2 = \frac{-d\tau^2 + \delta_{ij} dx^i dx^j}{H^2 \tau^2},
    \label{eq:dS_metric}
\end{equation}
where $\tau \in (-\infty, 0)$ is the conformal time and $H$ is the constant Hubble parameter.
The aether field $u^\mu$ is assumed to align with the cosmic time direction, respecting spatial translation and rotation symmetries. Normalizing it to satisfy the constraint $u^\mu u_\mu = -1$, we assume the ansatz~\cite{Carroll:2004ai}:
\begin{equation}
    u^\mu = - H \tau \delta^\mu_0, \quad \text{or equivalently} \quad u_\mu = \frac{1}{H\tau} \delta^0_\mu.
    \label{eq:u_ansatz}
\end{equation}

To evaluate the equations of motion, we first compute the covariant derivative of the aether field, $\nabla_\mu u_\nu = \partial_\mu u_\nu - \Gamma^\lambda_{\mu\nu} u_\lambda$. The non-vanishing Christoffel symbols for the metric \eqref{eq:dS_metric} are:
\begin{equation}
 \Gamma^\tau_{\tau\tau} = -\frac{1}{\tau}, \quad \Gamma^\tau_{ij} = -\frac{1}{\tau}\delta_{ij}, \quad \Gamma^i_{\tau j} = -\frac{1}{\tau}\delta^i_j.
\end{equation}

Substituting these into the definition of the covariant derivative yields the non-zero components:
\begin{equation}
    \nabla_\tau u_\tau = 0, \quad \nabla_i u_j = \frac{1}{H\tau^2} \delta_{ij}.
\end{equation}
These components can be covariantized into a compact tensor form proportional to the spatial projection tensor:
\begin{equation}
    \label{eq:nabla_u_dS}
    \nabla_\mu u_\nu = H (g_{\mu\nu} + u_\mu u_\nu).
\end{equation}
From this expression, it is evident that the aether flow is geodesic, as the acceleration vector vanishes identically:
\begin{equation}
    a^\mu \equiv u^\nu \nabla_\nu u^\mu = H u^\nu (g_{\nu\mu} + u_\nu u_\mu) = 0.
\end{equation}

Substituting this result into the definition of the aether energy-momentum tensor $T^{\mathrm{AE}}_{\mu\nu}$, terms involving acceleration and shear drop out. Since the configuration respects spatial isotropy, the energy-momentum tensor must take the form of a perfect fluid. However, due to the constraint $u^2=-1$ and the specific form of $\nabla_\mu u_\nu$, the anisotropic stress vanishes, and it reduces to a renormalization of the cosmological constant~\cite{Carroll:2004ai, Zlosnik:2006zu}:
\begin{equation}
    T_{\mu\nu}^{\mathrm{AE}} = \left( \frac{3\alpha}{2} M_{\mathrm{pl}}^2 H^2 \right) g_{\mu\nu},
\end{equation}
where the parameter $\alpha$ is defined by the coupling constants as $\alpha \equiv c_1 + 3c_2 + c_3$.

The Einstein equations, $G_{\mu\nu} + \Lambda g_{\mu\nu} = M_{\mathrm{pl}}^{-2} T_{\mu\nu}^{\mathrm{AE}}$, then yield the modified Friedmann equation:
\begin{equation}
    3 M_{\mathrm{pl}}^2 H^2 = \rho_\Lambda - \frac{3}{2} \alpha M_{\mathrm{pl}}^2 H^2,
\end{equation}
where we have defined the vacuum energy density associated with the bare cosmological constant as $\rho_\Lambda \equiv M_{\mathrm{pl}}^2 \Lambda$. Solving for the expansion rate $H$, we find:
\begin{equation}
    3 M_{\mathrm{pl}}^2 \left( 1 + \frac{\alpha}{2} \right) H^2 = \rho_{\Lambda}.
\end{equation}
This confirms that the Poincar\'e de Sitter metric is a consistent solution to the theory, provided the bare parameters satisfy this relation.

Finally, we characterize the symmetry-breaking pattern of the cosmological background described by \eqref{eq:dS_metric} and \eqref{eq:u_ansatz}. The background geometry possesses the full isometry group of de Sitter spacetime, $SO(1,4)$. A transformation generated by a Killing vector $\xi^\mu$ is a symmetry of the background configuration only if it preserves both the metric and the aether field:
\begin{equation}
    \mathcal{L}_\xi g_{\mu\nu} = 0 \quad \text{and} \quad \mathcal{L}_\xi u^\mu = 0.
\end{equation}
The isometries of the Poincar\'e patch are generated by spatial translations $P_i$, spatial rotations $M_{ij}$, dilatation $D$, and de Sitter boosts (or special conformal transformations) $K_i$.
Explicit calculation shows that the aether field $u^\mu = -H\tau \delta^\mu_0$ is invariant under translations, rotations, and dilatation~\cite{Green:2020ebl}:
\begin{equation}
    \mathcal{L}_{P_i} u^\mu = 0, \quad \mathcal{L}_{M_{ij}} u^\mu = 0, \quad \mathcal{L}_{D} u^\mu = 0.
\end{equation}
However, for the de Sitter boost generators $K_i$, the Lie derivative is non-vanishing:
\begin{equation}
    \mathcal{L}_{K_i} u^\mu \neq 0.
\end{equation}

Thus, the presence of the aether field spontaneously breaks the $SO(1,4)$ symmetry. The residual symmetry group is generated by spatial translations $P_i$, spatial rotations $M_{ij}$, and dilatation $D$.
Algebraically, this corresponds to the \textit{Similitude group} of $\mathbb{R}^3$, denoted as $\mathrm{Sim}(3)$. This group has the structure of a semi-direct product:
\begin{equation}
    G_{\mathrm{res}} \cong ISO(3) \rtimes \mathbb{R}_D,
\end{equation}
where $\mathbb{R}_D$ denotes the dilatation subgroup.
In the context of the holographic dual, this signifies that while the holographic Euclidean field theory retains scale invariance (dilatation), it does not exhibit full conformal invariance~\cite{Pajer:2020wxk}.

\section{Holographic Stress Tensor and Virial Current}

We consider the dual field theory defined on a flat three-dimensional background. In this setup, the distinction between scale invariance and full conformal invariance is encoded in the trace of the symmetric, conserved stress tensor $\mathcal{T}_{ij}$.
While conformal invariance requires the trace to vanish (potentially up to anomaly terms), scale invariance alone imposes a weaker condition: the trace is related to the divergence of a vector operator $\mathcal{J}_i$, known as the virial current~\cite{Polchinski:1987dy, Nakayama:2013is}:
\begin{align}
    \mathcal{T}^i_i = \partial^i \mathcal{J}_i .
\end{align}
The symmetry is enhanced to full conformal invariance if and only if the virial current itself is a total divergence of a local tensor $\mathcal{J}_i = \partial^j \mathcal{L}_{ji}$ or vanishes identically.

The goal of this section is to demonstrate this structure within the holographic framework of the Einstein-Aether theory. We derive the relationship between the trace of the holographic stress tensor and the divergence of the vector field components on the cosmological background. We show that the Hamiltonian constraint of the bulk Einstein equations relates the trace of the holographic stress tensor to the divergence of the subleading mode of the vector field, thereby identifying this mode as the virial current.

\subsection{Fefferman-Graham Expansions in dS}

We adopt the Fefferman-Graham gauge for the asymptotically de Sitter metric. Working in terms of the conformal time $\tau$ (with the future boundary located at $\tau \to 0$), the metric takes the form~\cite{Skenderis:2002wp}:
\begin{align}
    ds^2 = \frac{ -d\tau^2 + g_{ij}(\tau, x) dx^i dx^j }{{H^2 \tau^2}}.
\end{align}
The spatial metric component admits the following asymptotic expansion near $\tau \to 0$:
\begin{equation}
    g_{ij}(\tau, x) = g_{(0)ij}(x) + \tau^2 g_{(2)ij}(x) + \tau^3 g_{(3)ij}(x) + \cdots.
\end{equation}
Here, $g_{(0)ij}(x)$ represents the holographic metric on the spacelike slice $\mathcal{I}^+$. According to the standard holographic dictionary~\cite{Gubser:1998bc,Witten:1998qj,deHaro:2000vlm}, the coefficient $g_{(3)ij}(x)$ encodes the expectation value of the holographic stress tensor $\mathcal{T}_{ij}$. Specifically, the contraction $\mathrm{Tr}(g_{(3)}) \equiv g_{(0)}^{ij}g_{(3)ij}$ determines the trace of the holographic stress tensor.

We treat the bulk aether field $u_\mu$ as a vector field dual to a holographic vector operator of scaling dimension $\Delta = 2$. The covariant spatial components $u_i$ expand as:
\begin{equation}
    u_i(\tau, x) = v_{(0)i}(x) + \tau v_{(1)i}(x) + \cdots
    \label{eq:u_covariant},
\end{equation}
where $v_{(0)i}$ acts as the source, and $v_{(1)i}$ denotes the subleading response mode. The latter corresponds to the holographic vector operator $\mathcal{J}_i$, which we identify as the virial current in the subsequent analysis. For the linearized analysis presented in this section, the temporal component is fixed to its background value $u_\tau(\tau,x) \approx \frac{1}{H\tau}$, as corrections are of higher order due to the unit-norm constraint.

\subsection{Hamiltonian Constraint}

To obtain the operator relations such as $\mathcal{T}^i_i = \partial^i \mathcal{J}_i$, we need to derive the constraint equation from the bulk equations of motion. Such a condition arises from the Hamiltonian constraint, which relates the metric and matter field expansions near the boundary~\cite{Papadimitriou:2004ap, Skenderis:2002wp}.

The $(\tau, \tau)$ component of the Einstein equations, which constitutes the Hamiltonian constraint, is given by:
\begin{equation}
    G^\tau_\tau + \Lambda = M_{\mathrm{pl}}^{-2} T^\tau_\tau.
\end{equation}
It is important to note that the Hubble constant $H$ is determined not only by the bare cosmological constant $\Lambda$ but also by the background contribution from the aether energy-momentum tensor. We analyze this equation order by order in the conformal time $\tau$.

Let us begin with the left-hand side.
Using the Gauss-Codazzi relations adapted for the temporal coordinate $\tau$, the $(\tau, \tau)$ component of the Einstein tensor expands as:
\begin{equation}
    G^\tau_\tau = -3H^2 +2 H^2\mathrm{Tr}(g_{(2)}) \tau^2 + 3H^2 \mathrm{Tr}(g_{(3)}) \tau^3 + \mathcal{O}(\tau^4),
\end{equation}
Throughout the analysis, we assume $g_{(0)ij} = \delta_{ij}$.
Substituting this expansion into the field equation, we examine the contributions at each order.

At order $\tau^0$, the leading geometric term is $-3H^2$. This term is cancelled by the contributions from the cosmological constant and the energy-momentum tensor of the background aether field, consistent with the modified Friedmann equation derived in the previous section.
At order $\tau^2$, the terms provide the consistency condition that determines $g_{(2)}$ in terms of the boundary curvature $R[g_{(0)}]$ and aether sources, which we have turned off for simplicity. Our primary interest lies in the term of order $\tau^3$. The dynamical constraint appears at this order:
\begin{equation}
    3H^2 \mathrm{Tr}(g_{(3)}) \tau^3 = M_{\mathrm{pl}}^{-2} T^\tau_\tau \Big|_{\mathcal{O}(\tau^3)}.
    \label{eq:constraint_tau3}
\end{equation}

We now compute the right-hand side. Substituting the Fefferman-Graham expansions into the energy-momentum tensor $T^{\mathrm{AE}}_{\mu\nu}$ and linearizing around the background, we identify the contribution from the aether field fluctuations.
Assuming that the source terms vanish ($v_{(0)i} = 0$), the expansion yields:
\begin{equation}
    M_{\mathrm{pl}}^{-2} T^\tau_\tau \Big|_{\mathcal{O}(\tau^3)} = -\mathcal{K}(c_i)  H^2 \tau^3 (\partial^i v_{(1)i}) - H^2 \tau^3 \frac{3\alpha}{2} \mathrm{Tr}(g_{(3)}).
\end{equation}
Here, $\mathcal{K}(c_i) = \frac{2c_1 +3c_2 + c_3}{H}$ is a specific linear combination of the aether coupling constants, and $\alpha = c_1 + 3c_2 +c_3$ as defined before.
Equating the left-hand side and right-hand side, we obtain the constraint on the sub-leading modes:
\begin{equation}
   \left(1+\frac{\alpha}{2}\right) \mathrm{Tr}(g_{(3)}) = -\frac{\mathcal{K}(c_i)}{3} (\partial^i v_{(1)i}).
    \label{eq:Ttautau_final}
\end{equation}

Through the dS/CFT dictionary, $\mathrm{Tr}(g_{(3)})$ is proportional to the trace of the holographic stress tensor through the relation $\mathcal{T}_{ij} = \frac{3 M_{\mathrm{pl}}^2}{H^2} g_{(3)ij}$~\cite{deHaro:2000vlm}. This yields the Ward-Takahashi identity for the scale transformation:
\begin{equation}
   \langle \mathcal{T}^i_i \rangle = \langle \partial^i \mathcal{J}_i \rangle, \label{WI}
\end{equation}
where we have identified the virial current $\mathcal{J}^i$ with the subleading mode of the aether field:\footnote{Note that this relation does not say anything about the transverse part of $\mathcal{J}_i$ or $v_{(1)i}$.}
\begin{equation}
    \partial^i\mathcal{J}_i = -\frac{\mathcal{K}(c_i)}{H^2  \left(1+\frac{\alpha}{2}\right) } \partial^i v_{(1)i}.
\end{equation}
The proportionality constant depends on $c_i$ and $H$. This result confirms that the non-vanishing divergence of the vector field in the bulk---arising from the boost-breaking sector---sources the trace of the holographic stress tensor, a hallmark of a theory that is scale invariant but not conformal invariant.

\subsection{Two-point functions and power spectrum}
In order to show that the divergence of the holographic virial current operator is non-vanishing in this setup, it is sufficient to study its two-point function. Let us consider the spatial fluctuation of the future boundary metric in the momentum space $ \delta g_{ij}(\mathbf{k})$.
The holographic dictionary states that the wave functional of the universe takes the form~\cite{Maldacena:2002vr}:
\begin{align}
    \Psi[\delta g_{ij}] = \exp\left( - \frac{1}{2} \int \frac{d^3k}{(2\pi)^3}\frac{d^3p}{(2\pi)^3} \, \delta g_{ij}(\mathbf{k}) \langle \mathcal{T}^{ij}(\mathbf{k}) \mathcal{T}^{kl}(\mathbf{p}) \rangle \delta g_{kl}(\mathbf{p}) \right)
\end{align}
at the quadratic order in the metric perturbation.
Here, $\mathcal{T}_{ij}$ is the stress tensor of the (hypothetical) holographic Euclidean field theory.

Without specifying the details of the holographic Euclidean field theory, we can constrain the form of the two-point functions of the holographic stress tensor based on symmetry arguments.
We can always decompose the correlation function of the holographic stress tensor into spin-2 (tensor) and spin-0 (scalar) contributions using the standard projection operators:
\begin{align}
    \langle\langle \mathcal{T}_{ij}(\mathbf{k}) \mathcal{T}_{kl}(-\mathbf{k}) \rangle\rangle = A(k) \Pi_{ij,kl}^{(2)} + B(k) \Pi_{ij,kl}^{(0)}. \label{TPF}
\end{align}
Here, the double bracket $\langle\langle \cdots \rangle\rangle$ means the delta function $(2\pi)^3 \delta(\mathbf{k}_1 + \mathbf{k}_2)$ is omitted, and $\Pi^{(2)}$ and $\Pi^{(0)}$ are the transverse-traceless and scalar projectors, defined respectively as:
\begin{align}
    \Pi_{ij,kl}^{(2)} &= \frac{1}{2} \left( P_{ik} P_{jl} + P_{il} P_{jk} - P_{ij} P_{kl} \right), \\
    \Pi_{ij,kl}^{(0)} &= \frac{1}{2} P_{ij} P_{kl},
\end{align}
with the transverse projector $P_{ij} \equiv \delta_{ij} - k_i k_j/k^2$. Note that this decomposition assumes the conservation of the stress tensor ($\partial^i \mathcal{T}_{ij} = 0$) and spatial isotropy, which are guaranteed by the holographic Ward-Takahashi identities.

It is important to comment on the locality of this decomposition. From the viewpoint of the dual field theory, the decomposition into $A(k)$ and $B(k)$ involves projection operators that are non-local in position space. On the other hand, in the bulk computation, at the linear perturbation theory, the spin two mode and spin zero mode are decoupled without causing any non-locality issue.

We will show that $A(k)$ and $B(k)$ have scale-invariant form, but the form of the power spectrum alone does not tell if the scalar mode is associated with the virial current. In particular, the crucial question if the virial current has the structure $\mathcal{J}_i = \partial^i \mathcal{L}_{ij}$ so that it has hidden conformal invariance cannot be addressed.\footnote{See \cite{Kawai:2014vxa} for an example of holographic cosmology where this ambiguity of the stress tensor was discussed.} In this sense, the study of the power spectrum is complementary to the discussion in the previous subsection.

The holographic dictionary relates $A(k)$ and $B(k)$ in \eqref{TPF} to the tensor and scalar power spectra of the gravitational wave:
\begin{align}
    P_\gamma(k) &= \frac{1}{2 \mathrm{Re} A(k)}, \\
    P_\zeta(k) &= \frac{1}{8 \mathrm{Re} B(k)}.
\end{align}
This explicitly shows that the scalar power spectrum is determined by the scalar component $B(k)$ of the holographic stress tensor correlation.

If the holographic field theory were a conformal field theory, the trace of the stress tensor would vanish ($ \mathcal{T}^i_i = 0$), implying $B(k)=0$ in this decomposition. In such a case, there would be no source for the scalar gravitational wave. Thus, our goal is to show that the scalar power spectrum is finite even in the de Sitter background in the Einstein-Aether theory, indicating non-trivial conformal symmetry breaking.

The computation of the power spectra in Einstein-Aether theory is conceptually straightforward but technically involved. Therefore, we quote the standard results found in the literature~\cite{Lim:2004js,Jacobson:2007veq, Armendariz-Picon:2010aer}.\footnote{See also \cite{Dong:2026xab} for the propagation in a more general background.} We use the notation $c_{ij} \equiv c_i + c_j$ and $c_{123} \equiv c_1+c_2+c_3$, which is commonly used there.

Let us start with the transverse-traceless metric perturbation. We can quantize the tensor mode in the standard manner from the effective action $S =\int dt \frac{a^2}{2Z_T} \left( \dot{h}^2 - c_{T}^2 k^2 h^2 \right)$,
and obtain the tensor power spectrum $P_h(k)$:
\begin{equation}
    P_{h}(k) = \frac{ Z_T}{ 2c^3_T} \frac{H^2}{k^3},
    \label{eq:tensor_spectrum_exact}
\end{equation}
where the amplitude and the tensor sound speed are given by 
\begin{align} 
Z_T &= (1-c_{13})^{-1}{M_\mathrm{pl}}^{-2} \\ 
c_T^2 &= \frac{1}{1 - c_{13}}.
\end{align}

In a similar manner, the scalar power spectrum is (approximately) given by
\begin{align}
    P_{\zeta}(k) &= \frac{Z_A}{2 c_s^3} \frac{H^2}{k^3},
\end{align}
where the amplitude and the sound speed of the scalar mode are given by
\begin{align}
Z_A &= \frac{c_{123} c_T^2}{2+c_{13}+3c_2} \frac{M_{\mathrm{pl}}^{-2}}{2}   \\
    c_s^2 &= \frac{c_{123}(2-c_{14})}{c_{14}(1-c_{13})(2+c_{13}+3c_2)}.
\end{align}
See, e.g.,~\cite{Oost:2018tcv}. Here, we have assumed the sub-horizon limit and standard Bunch-Davies vacuum.

We immediately see that both power spectra are scale invariant ($P(k) \propto k^{-3}$). However, the non-vanishing result for $P_\zeta$ implies that the dual field theory is not conformal invariant unless the scalar part is removed by the field redefinition corresponding to the situation $\mathcal{J}_i = \partial^j \mathcal{L}_{ij}$ (see \cite{Kawai:2014vxa}). Again, we emphasize that this possibility cannot be ignored by only looking at the shape of the scalar power spectra, but we assume this is not the case from the discussions in the previous subsection.\footnote{This may happen when the aether field can be written as $v_\mu = \partial_\mu \varphi$ with a local bulk field $\varphi$.}

Finally, we briefly comment on the vector mode. The vector mode corresponds to the transverse part of the virial current operator. In the cosmological computation, the vector mode decouples from the tensor and scalar modes at the linearized level. Similar to the scalar case, the decomposition of the transverse part and the non-transverse part of the virial current is a non-local manipulation within the holographic field theory.

\subsection{Three-point functions and non-Gaussianity}

The presence of the background aether field modifies matter interactions in both the kinetic and potential terms. This gives rise to three-point functions (non-Gaussianity) that are scale-invariant but fail to preserve full conformal invariance. While a complete analysis of the bispectra in the Einstein-Aether theory is technically demanding, we illustrate the key features by studying a probe scalar field coupled to the background aether.

The salient feature of boost symmetry breaking is that the propagation speed of scalar waves is not fixed to unity (the speed of light). This deviation leads to a violation of the full conformal symmetry in cosmological correlation functions. To isolate this effect, we examine the three-point functions of scalars with distinct propagation speeds.

We begin with the action for a scalar field $\phi$. The coupling to the aether field $u^\mu$ introduces a modification to the kinetic term:
\begin{align}
    S_\phi = -\int d^4x \sqrt{-g}\left( \frac{1}{2} g^{\mu\nu}\partial_\mu \phi \partial_\nu \phi + \frac{1}{2} \gamma (u^\mu \partial_\mu \phi)^2 + \frac{1}{2} m^2 \phi^2 + V_{\mathrm{int}} \right).
\end{align}
Assuming the background aether alignment $u^\mu = - H \tau \delta_0^\mu$ in the de Sitter background, the term $(u^\mu \partial_\mu \phi)^2$ modifies the time derivative component. This effectively renormalizes the sound speed of the scalar field to $c_s^2 = (1 - \gamma )^{-1}$.

To demonstrate the impact on cosmological correlators, we first consider an illustrative case: three scalar fields with ``conformal mass'' $m^2 = 2H^2$ (corresponding to scale dimension $\Delta = 2$) but with distinct sound speeds $c_1, c_2, c_3$. We assume they interact via a non-derivative interaction $V_{\mathrm{int}} = \lambda \phi_1 \phi_2 \phi_3$.
For a conformally coupled scalar in de Sitter spacetime, the mode function takes the simple form $u_k(\tau) = \frac{H\tau}{\sqrt{2 c_i k}} e^{-i c_i k \tau}$.
The computation of the three-point function follows the standard in-in formalism. Up to overall numerical coefficients, the result scales as:
\begin{align}
    \langle \langle O_1(k_1) O_2(k_2) O_3(k_3) \rangle \rangle_{\lambda} \propto \lambda \log (c_1 k_1 + c_2 k_2 + c_3 k_3).
\end{align}
This three-point function is conformal invariant only if the speeds are degenerate, $c_1 = c_2 = c_3$. The logarithmic dependence signifies the presence of a Weyl anomaly, a known feature of the dual field theory~\cite{Nakayama:2013wda}. When the velocities differ, the argument of the logarithm breaks the permutation symmetry associated with the full conformal group. To see this, recall that the Polyakov formula of the conformal three-point function is permutation invariant if $\Delta_1 = \Delta_2 = \Delta_3$.

Next, we consider the case of massless scalar fields ($\Delta = 3$) with interaction structures induced by the aether. The aether field allows for interactions involving the preferred time direction, denoted by the operator $\mathcal{D}_u \phi \equiv u^\mu \partial_\mu \phi$. We classify the interactions by the number of time derivatives:
\begin{itemize}
    \item $V_1 = \lambda_1 (\mathcal{D}_u \phi_1)\phi_2 \phi_3$
    \item $V_2 = \lambda_2 (\mathcal{D}_u \phi_1)(\mathcal{D}_u \phi_2) \phi_3$
    \item $V_3 = \lambda_3 (\mathcal{D}_u \phi_1)(\mathcal{D}_u \phi_2)(\mathcal{D}_u \phi_3)$
\end{itemize}
The case $V_1$ has been addressed in the context of AdS holography~\cite{Skenderis:2002wp}. Here we focus on $V_2$ and $V_3$, which are characteristic of boost-breaking theories.

For massless scalars in de Sitter space with sound speed $c$, the mode function is $u_k(\tau) = \frac{H}{\sqrt{2 c^3 k^3}}(1+i c k \tau)e^{-i c k \tau}$. 
For the interaction $V_2 = \lambda_2 \dot{\phi}_1 \dot{\phi}_2 \phi_3$, the resulting three-point function is givne by
\begin{align}
    \langle\langle O_1(k_1) O_2(k_2) O_3(k_3) \rangle\rangle_{\lambda_2} \propto  \lambda_2  k_1^2 k_2^2 \left( \frac{1}{E} + \frac{c_3 k_3}{E^2} \right),
\end{align}
where $E \equiv c_1 k_1 + c_2 k_2 + c_3 k_3$ is the total energy. Again, it is easy to see that the three-point function is not conformal invariant because it is not permutation invariant.

%This shape differs from the simple $1/E$ pole found in conformal scalars; the interference between the scale factor and the massless mode functions generates a double pole $1/E^2$, modifying the shape of non-Gaussianity.

Similarly, for the interaction $V_3 = \lambda_3\dot{\phi}^3$, the three-point function is given by:
\begin{align}
    \langle \langle O_1(k_1) O_2(k_2) O_3(k_3) \rangle\rangle_{\lambda_3} \propto \lambda_3 \frac{k_1^2 k_2^2 k_3^2}{E^3}.
\end{align}
This result exhibits a generalized equilateral shape. The distinct sound speeds $c_i$ appearing in the denominator $E^3$ explicitly demonstrate the violation of de Sitter isometry due to the aether background and hence the violation of the conformal invariance.

Let us add that the conformal invariance of the three-point functions from $V_1$, $V_2$, and $V_3$ is violated even if $c_1=c_2=c_3$ because the interaction vertex itself is not de Sitter invariant. If we used the vertex $V_0 = \lambda_0 \phi_1 \phi_2 \phi_3$, the three-point functions would be conformal invariant, but in the actual computation, we should face a severe IR divergence of the massless scalar propagation in de Sitter spacetime. 

\section{An Example of Scale but Non-Conformal Field Theory}

Here, we take the simplest example of a scale-invariant but non-conformal field theory: Maxwell theory in three dimensions. We verify the non-vanishing trace of the stress tensor and compute its two-point and three-point functions, which should be compared with the perturbative holographic computation discussed in the previous section. Unlike in the conformal invariant cases, it is highly non-trivial because the symmetry does not fix the three-point functions. Indeed, we will see that it is significantly different from what we obtained in the previous section.

The Euclidean action is given by:
\begin{align}
 S =  \int d^3x \left( -\frac{1}{4} F_{ij} F^{ij} \right).
\end{align}
In three dimensions, the coupling constant is dimensionful, but the free theory is scale invariant. The symmetric stress tensor and its trace are given by:
\begin{align}
\mathcal{T}_{ij} &= F_{ik} F_{j}^{\  k} -\frac{1}{4} \delta_{ij} F_{kl} F^{kl}, \\ 
\mathcal{T}^{i}_{i} &= \frac{1}{4} F_{kl} F^{kl} = \frac{1}{2} \partial_k (A_l F^{kl}). 
\end{align}
Here, the trace is non-zero, satisfying the relation $\mathcal{T}^i_i = \partial^i \mathcal{J}_i$ with the virial current $\mathcal{J}_i = \frac{1}{2} A^l F_{il}$. Note that $\mathcal{J}_i$ is not gauge invariant, which is a crucial feature of this theory~\cite{El-Showk:2011xbs, Jackiw:2011vz}.

We can straightforwardly compute the correlation functions of the trace of the stress tensor $\mathcal{T}^i_i$. Since the theory is free and Gaussian, the correlators are obtained via Wick contractions.
The two-point function and the three-point function in momentum space are non-zero and take the following forms (up to overall normalization constants):
\begin{align}
 \langle\langle \mathcal{T}_i^i(\mathbf{k}) \mathcal{T}_{i}^i(-\mathbf{k}) \rangle\rangle &\sim k^3, \\
 \langle\langle \mathcal{T}_{i}^i(\mathbf{k}_1) \mathcal{T}_i^i(\mathbf{k}_2) \mathcal{T}_i^i(\mathbf{k}_3) \rangle\rangle  &\sim  2k_1 k_2 k_3 - (k_1 + k_2 + k_3)(k_1^2 + k_2^2 + k_3^2) + 4(k_1^3 + k_2^3 + k_3^3).
\end{align}
This three-point function in coordinate space was computed in \cite{El-Showk:2011xbs} in general dimensions while the momentum space results in $d=3$ can be found e.g. in \cite{McFadden:2010vh}. 
It is worth noting that for the three-point function, the local polynomial terms (contact terms and semi-contact terms) may be subject to regularization scheme dependence, but the non-local structure is robust

At the technical level, it is curious to note that the three-point function does {\it not} appear to be proportional to $k_1^i k_2^j k_3^l$ although we are supposed to compute the correlation functions of the divergence of the virial current $\mathcal{T}_i^i = \partial^i \mathcal{J}_i$. A non-trivial cancellation occurs with the denominator, i.e., $(k_1^2 k_2^2 k_3^2)/(k_1 k_2 k_3) = k_1 k_2 k_3$. Such a cancellation makes it difficult to see if the operator is a derivative of some other operators or not.

These non-vanishing trace correlation functions explicitly demonstrate that the theory breaks conformal invariance despite being scale invariant.
Comparing this result with the bulk computation in the previous section, however, reveals a significant difference. The three-point functions in the Einstein-Aether theory typically contain ``energy poles" of the form $1/E^n$ (where $E = k_1+k_2+k_3$), which arise from the bulk propagation of massless modes interacting at a point. In contrast, the free Maxwell result consists of a sum of cubic terms and lacks the $1/E$ pole structure.
This discrepancy suggests that the cosmology arising from the dual Maxwell theory is likely non-local or involves higher-spin fields, rather than the simple effective field theory described by the Einstein-Aether action.\footnote{There is a similar bulk-point singularity question in AdS holography ~\cite{Gary:2009ae,Heemskerk:2009pn,Penedones:2010ue}.}

Furthermore, the gauge dependence of the virial current $\mathcal{J}_i$ implies that there is no gauge-invariant local operator in the dual field theory to source the bulk vector field $v_i$. Thus, a direct mapping to the standard Einstein-Aether bulk is not expected for Maxwell theory.
To find a precise holographic dual described by Einstein-Aether gravity, one should look for systems where the virial current is a well-defined physical operator. Promising candidates include the theory of elasticity~\cite{Riva:2005gd}, membranes~\cite{Mauri:2021ili}, dipolar magnets~\cite{Gimenez-Grau:2023lpz}, or random spin systems~\cite{Nakayama:2024jwq}, where the virial current has a clear physical interpretation and can be identified with a physical bulk vector field.

\section{Discussion and Summary}

In this paper, we have investigated the holographic structure of the Einstein-Aether theory, with a particular focus on the distinction between scale invariance and full conformal invariance.
Our primary theoretical finding is the identification of the subleading mode of the bulk aether field with the virial current operator $\mathcal{J}_i$ in the dual field theory.
 Through the holographic dictionary and the analysis of the Hamiltonian constraint, we derived the operator relation $\mathcal{T}^i_i = \partial^i \mathcal{J}_i$. This Ward-Takahashi identity explicitly demonstrates that the non-vanishing divergence of the virial current---sourced by the bulk aether mode---breaks special conformal transformations while preserving scale invariance.

We further explored the observational consequences of this symmetry-breaking pattern in the context of primordial cosmology. The presence of the aether field induces a non-vanishing scalar power spectrum $P_\zeta$. In a standard conformal field theory, the trace of the stress tensor vanishes, leaving no source for the scalar mode; thus, a finite $P_\zeta$ is a direct signature of conformal symmetry breaking.
Moreover, our analysis of the bispectrum (three-point function) revealed that the breaking of boost invariance manifests as a dependence on the individual sound speeds $c_i$ in the pole structure, taking forms such as $(c_1 k_1 + c_2 k_2 + c_3 k_3)^{-n}$. This distinctive ``speed-dependent" non-Gaussianity serves as a fingerprint of the Einstein-Aether theory, distinguishing it from standard single-clock inflation or conformal invariant holographic scenarios where the speed of sound is unique.

We also compared our holographic setup with a well-known example of a scale-invariant but non-conformal field theory: Maxwell theory in three dimensions.
While this theory shares the property $\mathcal{T}^i_i \neq 0$, it presents a subtlety regarding holographic duality. In three-dimensional Maxwell theory, the virial current is not gauge-invariant, suggesting that a naive bulk dual with a vector field (which typically couples to a gauge-invariant operator) may not fully capture its physics.
Instead, our Einstein-Aether setup may be more naturally dual to condensed matter systems where the virial current is a well-defined, gauge-invariant physical observable, such as in the theory of elasticity, dipolar magnets, or random spin systems.

There are several promising directions for future research.
First, the non-Gaussianity shapes derived here are relevant to the program of ``Cosmological Collider Physics."~\cite{Chen:2009zp,Noumi:2012vr,Arkani-Hamed:2015bza} The massive spin-1 or spin-0 modes in the aether sector could leave oscillatory signatures in the squeezed limit of the bispectrum, providing a unique probe of the bulk mass spectrum and symmetry-breaking pattern.
Second, extending this analysis to theories with Ho\v{r}ava-Lifshitz scaling would be valuable. Since Einstein-Aether theory can be viewed as the low-energy limit of Ho\v{r}ava gravity, understanding the holographic renormalization group flow between these regimes would provide deeper insights into the emergence of spacetime and Lorentz invariance. 
Third, it would be interesting to revisit the effective field theory of vector-tensor theories~\cite{Aoki:2021wew,Bertucci:2024qzt} from the viewpoint of the distinction between scale and conformal invariance.

Finally, an ultimate question regarding fundamental symmetries remains: Is scale invariance without conformal invariance a natural feature of fundamental physics, or does nature prefer the enhanced symmetry of the conformal group? Is there a theoretical principle that explains the tight observational bounds~\cite{Yagi:2013qpa,Oost:2018tcv,Gupta:2021vdj,Adam:2021vsk,Schumacher:2023cxh} on the aether parameters that violate boost symmetry?
To boost or not to boost, that is the question.

\section*{Acknowledgements}
The author thanks D.~Dorigoni and M,~Honda for inviting me to Japan-UK Workshop on Quantum Gravity in Kobe, where he gave a talk with the same title (and failed British joke). He also thanks G.~Pimentel for the discussions and the best chocolate in Italy. The work by YN is in part supported by JSPS KAKENHI Grant Number 21K03581.

\bibliographystyle{JHEP}
\bibliography{main}

\end{document}